\documentclass[conference]{IEEEtran}
\IEEEoverridecommandlockouts

\usepackage{cite}
\usepackage{amsmath,amssymb,amsfonts}
\usepackage{algorithmic}
\usepackage{graphicx}
\usepackage{textcomp}
\usepackage{booktabs}
\usepackage{xcolor}
\usepackage[hyphens]{url}
\usepackage[T1]{fontenc}

\makeatletter
\newcommand{\newlineauthors}{%
  \end{@IEEEauthorhalign}
  \hfill\mbox{}\par
  \mbox{}\hfill
  \begin{@IEEEauthorhalign}
}
\makeatother

\begin{document}

\title{Silent on Code? A Computing Lens on Generative AI Guidance at Canadian Polytechnics and Research Universities\\
\thanks{This research was supported in part by the Social Sciences and Humanities Research Council of Canada (SSHRC) under grants \mbox{430-2026-00030} (PI: Marco Ho) and \mbox{435-2026-1299} (PI: Michael Lin).
}}

\author{
\IEEEauthorblockN{Marco Ho}
\IEEEauthorblockA{\textit{School of Computing \& Academic Studies} \\
\textit{British Columbia Institute of Technology}\\
Burnaby, BC, Canada \\
marco\_ho@bcit.ca}
\and
\IEEEauthorblockN{Michael Pin-Chuan Lin}
\IEEEauthorblockA{\textit{Faculty of Education} \\
\textit{Mount Saint Vincent University}\\
Halifax, NS, Canada \\
michael.lin@msvu.ca}
\and
\IEEEauthorblockN{Rishva Riteshkumar Amin}
\IEEEauthorblockA{\textit{Olsen College of Engineering \& Science} \\
\textit{Fairleigh Dickinson University}\\
Vancouver, BC, Canada \\
r.amin@student.fdu.edu}
 \and[\hfill\mbox{}\par\mbox{}\hfill]
\newlineauthors
\IEEEauthorblockN{Daniel Chang}
\IEEEauthorblockA{\textit{Faculty of Humanities \& Social Sciences} \\
\textit{Athabasca University}\\
Athabasca, AB, Canada \\
dchang@athabascau.ca}
\and
\IEEEauthorblockN{Jeeho Ryoo}
\IEEEauthorblockA{\textit{Olsen College of Engineering \& Science} \\
\textit{Fairleigh Dickinson University}\\
Vancouver, BC, Canada \\
j.ryoo@fdu.edu}
}
\maketitle

\begin{abstract}
Generative artificial intelligence (GenAI) tools now write, explain, and debug code, which changes how computing students learn and how their work is assessed. Instructors, however, act within institutional guidance written largely with essays in mind. We report a document analysis of the public, institution-level GenAI guidance of the 13 members of Polytechnics Canada and the 15 members of the U15 Group of Canadian Research Universities, retrieved in October 2026. Using a 27-item codebook spanning governance, integrity and assessment, data and legal issues, values, AI literacy, and a computing lens, we find that both sectors leave course-level decisions to instructors, although polytechnics more often prohibit use unless the instructor permits it (8 of 13 vs.\ 4 of 14), and converge on disclosure, links to academic integrity policy, and, in most institutions, the expectation that programs adapt rules to their discipline. Polytechnics frame GenAI as workforce preparation substantively far more often (7 of 13 vs.\ 0 of 14), whereas universities more often address assessment redesign. Twelve of 27 institutions with retrievable guidance mention code, but almost always as one output type among many; none names an AI coding assistant or provides substantive code-specific guidance on attribution, assessment, or licensing. Institution-level guidance is thus not silent on code but leaves programming-specific questions largely to programs and instructors. We propose five elements of discipline-level guidance for computing and release our codebook and coded dataset.
\end{abstract}

\begin{IEEEkeywords}
generative AI, institutional policy, academic integrity, computing education, polytechnics, document analysis
\end{IEEEkeywords}

\section{Introduction}

Large language models (LLMs) can solve many introductory programming exercises and explain or repair code on demand \cite{becker2023programming, denny2024computing}, and AI coding assistants are embedded in the professional tools that computing graduates will use. Computing educators have therefore debated whether to ban, tolerate, or integrate these tools, and how to adapt assessment \cite{lau2023ban, prather2023robots}.

Most post-secondary institutions responded with policies, position statements, or guidelines. Analyses of this guidance have examined highly ranked universities worldwide \cite{moorhouse2023generative}, United States research universities \cite{mcdonald2025generative}, and Canadian universities \cite{nagpal2024policies, marcel2024examining}; a study of Ontario institutions analyzed academic integrity policies that reference AI, including those of colleges \cite{miron2025policy}. Recent United States work has begun to examine generative AI (GenAI) rules in computing course syllabi and their relation to institutional guidance \cite{ali2025syllabi, ganguly2026comparative}. Two gaps remain. First, to our knowledge, no study has examined the GenAI guidance of Canadian polytechnics as a sector at the national level, even though these applied, industry-aligned institutions served 472,900 for-credit students in 2023--24 and frame their mission around building workforce-ready graduates \cite{polytechnicscanada2026work}. Second, to our knowledge, no study has coded institution-level guidance for programming-specific content, such as how generated code should be attributed, which coding assistants are permitted under privacy and intellectual property (IP) rules, how the licensing and security of generated code are addressed, and how programming skill is assessed when code can be generated automatically.

We address both gaps with a comparative document analysis of 28 Canadian institutions (27 with retrievable guidance) and three research questions.
\textbf{RQ1 (Governance):} How do institutions govern student use of GenAI, and who decides at the course level?
\textbf{RQ2 (Themes):} Which concerns does the guidance address, and how do polytechnics and research universities differ?
\textbf{RQ3 (Computing lens):} To what extent does institution-level guidance address programming-specific issues?

\section{Related Work}

\subsection{GenAI in Computing Education}
A working group of computing educators documented risks such as student over-reliance and new forms of academic misconduct~\cite{prather2023robots}, and one technical response has been to build assistants with guardrails that withhold complete solutions~\cite{liffiton2024codehelp}. In United States computing courses, most syllabus rules on GenAI appear within academic integrity statements and restrict its use~\cite{ali2025syllabi}, and course-level guidance tends to be more cautious than institution-level guidance~\cite{ganguly2026comparative}.

\subsection{Institutional GenAI Guidance} 
Analyses of top-ranked universities worldwide and of United States institutions found that most permitted GenAI use at the instructor's discretion and emphasized academic integrity, assessment redesign, and acknowledgment of use \cite{moorhouse2023generative, mcdonald2025generative}; the latter also warned that this can burden instructors \cite{mcdonald2025generative}. In Canada, analyses of leading universities found a shared emphasis on transparency and instructor autonomy but differing stances on AI detection and citation \cite{nagpal2024policies, marcel2024examining}, and college integrity policies in Ontario, Alberta, and Manitoba have been analyzed \cite{miron2025policy, stoesz2019ontario, eaton2023colleges}, though not the full range of GenAI guidance. International frameworks such as UNESCO's add expectations around equity, privacy, and capacity building \cite{unesco2023guidance}.

\section{Method}

\subsection{Sampling Frame and Documents}
We used qualitative document analysis \cite{bowen2009document} with a deductive codebook. Two externally defined groups of institutions were compared: the 13 members of Polytechnics Canada (Group P) \cite{polytechnicscanada2026work}, which includes two British Columbia (BC) institutions, the British Columbia Institute of Technology (BCIT) and Kwantlen Polytechnic University (KPU); and the 15 members of the U15 Group of Canadian Research Universities (Group U) \cite{u15members}. 

Because universities usually publish more guidance, one scope rule kept the material comparable: up to four public, institution-level documents on GenAI in teaching and learning were collected per institution, in priority order: (1) formal policies or procedures that mention GenAI; (2) position statements, principles, or official guidelines; (3) guidance from the central teaching centre, including sample syllabus statements; and (4) central guidance for students. News releases, individual course syllabi, faculty or department pages, research-only and administrative guidance, and pages behind a login were excluded. Documents were located with the same search template for every institution; targeted re-checks of institutions with few search results located documents that explicitly name AI at the Northern Alberta Institute of Technology (NAIT) and Saskatchewan Polytechnic, and a bilingual search located further University of Ottawa documents. All documents were retrieved on October 1, 2026. For one university, the Universit\'e de Montr\'eal, central guidance was identified through third-party documents but could not be retrieved. This institution is therefore excluded, leaving Group U with $n = 14$.

\subsection{Codebook and Coding}
The codebook contains 27 items grouped into six families (Tables~\ref{tab:ordinal} and~\ref{tab:categorial}). Ordinal items are coded on a three-point scale: 0 (absent), 1 (mentioned without actionable guidance), or 2 (substantive guidance, defined as a directive, a procedure, a worked example, or at least two sentences of specific guidance). Two computing items are coded present/absent only. Categorical items record three features: course-level decision locus (instructor decides, no default; prohibited unless instructor permits it; or not specified), the policy instrument (formal policy; guidelines only; or none found), and the stance on AI detectors (discouraged or unsupported; permitted with caution; or not mentioned). The computing family asks whether the guidance mentions code at all, describes use cases involving code, names an AI coding assistant, sets code-specific attribution rules, gives programming-specific assessment guidance, or addresses the licensing or security of generated code. General-purpose chat assistants, such as Microsoft Copilot, do not count as coding assistants. During piloting and coding, seven decision rules were logged. For example, content appearing only in optional sample syllabus statements is capped at level 1, and an institution's codes may be supported only by that institution's own documents. After the first six institutions had been coded, one inductive item (adaptation of guidance at the program or discipline level) was added, with earlier institutions recoded.

\begin{table}[t]
\caption{Ordinal Items from Polytechnics and Universities}
\label{tab:ordinal}
\centering
\footnotesize
\setlength{\tabcolsep}{4pt}
\begin{tabular}{@{}llcc@{}}
\toprule
Family & Item & P ($n=13$) & U ($n=14$) \\
\midrule
Governance & Syllabus statements & 12 (7) & 14 (11) \\
 & Link to integrity policy & 13 (10) & 14 (11) \\
 & Program/discipline adaptation & 8 (5) & 9 (4) \\
Integrity & Disclosure & 12 (10) & 14 (11) \\
 & Citation format & 11 (3) & 13 (3) \\
 & Assessment redesign & 7 (4) & 14 (8) \\
 & Process evidence & 8 (3) & 10 (3) \\
Data/legal & Privacy & 11 (9) & 13 (11) \\
 & Approved/restricted tools & 10 (9) & 13 (8) \\
 & Student work as AI input & 8 (5) & 6 (5) \\
 & Copyright/IP & 10 (6) & 13 (7) \\
Values & Equitable access & 8 (3) & 8 (8) \\
 & Accessibility & 6 (1) & 4 (3) \\
 & Environmental impact & 5 (2) & 9 (0) \\
 & Indigenous knowledge/data & 4 (3) & 1 (1) \\
 & Accuracy/bias & 13 (10) & 13 (10) \\
Literacy & Student AI literacy & 12 (8) & 14 (12) \\
 & Workforce framing & 9 (7) & 8 (0) \\
Computing & Code mentioned at all & 6 & 6 \\
 & Code use cases & 2 (0) & 2 (1) \\
 & Coding assistant named & 0 & 0 \\
 & Code attribution & 0 (0) & 1 (0) \\
 & Programming assessment & 1 (0) & 0 (0) \\
 & Code licensing/security & 1 (0) & 0 (0) \\
\bottomrule
\multicolumn{4}{@{}p{7.9cm}@{}}{\scriptsize Number of institutions with the item present (level 1 or 2); substantive (level 2) counts in parentheses. Code mentioned at all and coding assistant named are coded present/absent only.}
\end{tabular}
\end{table}

Document retrieval and coding were performed by an AI system (Claude Opus 5.5, Anthropic) applying the codebook and the logged decision rules under the authors' direction. All source documents and the manuscript were reviewed by the authors. Since a single coder applied the codebook, inter-rater reliability is not reported. A second pass re-read in full the core documents of a random 30\% of institutions (8 of the 25 then retrieved; fixed seed): agreement with the first pass was 88.4\% (191 of 216 codes; linear-weighted $\kappa = 0.86$ for the 0/1/2 items). Because the same system coded both passes without blinding, this shows consistency, not validity. As 21 of the 22 changes to ordinal items were increases, indicating that search extracts undercount guidance, the principal documents of all remaining institutions were then read in full. Overall, at least one core document was read in full for 25 of the 27 institutions (all but the University of British Columbia (UBC), whose principal page was read in part, and the University of Ottawa; 33 of 87 documents), and all codes resting on search extracts, including zeros, are flagged in the dataset. Given group sizes of 13 and 14, sector differences are reported descriptively, without significance tests.

The protocol and codebook, the list of documents with retrieval dates, the coded dataset (labelled as AI-coded, with an evidence note for every nonzero code), and the analysis scripts are openly available \cite{ho2026dataset}.

\section{Results}

\begin{figure*}[t]
\centering
\includegraphics[width=0.94\textwidth]{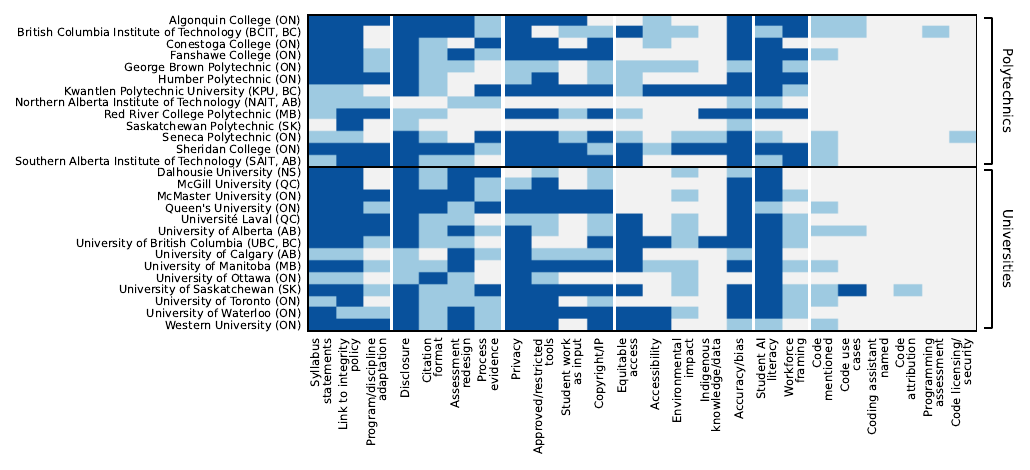}
\caption{Institution-by-item coding: (2) dark, substantive; (1) light, mentioned; (0) grey, absent. The computing block (right) is sparse in both sectors.}
\label{fig:heat}
\end{figure*}

\begin{table}[t]
\caption{Categorical Items from Polytechnics and Universities}
\label{tab:categorial}
\centering
\footnotesize
\setlength{\tabcolsep}{4pt}
\begin{tabular}{@{}llcc@{}}
\toprule
Item & Category & P ($n=13$) & U ($n=14$) \\
\midrule
Decision locus & Instructor decides, no default & 4 & 9 \\
 & Prohibited unless permitted & 8 & 4 \\
 & Not specified & 1 & 1 \\
Instrument & Formal policy names AI & 4 & 4 \\
 & Guidelines only & 9 & 10 \\
AI detectors & Discouraged or unsupported & 3 & 7 \\
 & Permitted with caution & 3 & 0 \\
 & Not mentioned & 7 & 7 \\
\bottomrule
\multicolumn{4}{@{}p{7.9cm}@{}}{\scriptsize Number of institutions; each institution falls into exactly one category per item.}
\end{tabular}
\end{table}

Paired counts in this section are reported as polytechnics vs. universities (out of 13 and 14 institutions, respectively).

\subsection{Governance (RQ1)}
Both sectors place course-level decisions with instructors, but polytechnics more often state a default of no use unless the instructor permits it (8 vs.\ 4; Table~\ref{tab:categorial}); among universities, these are Toronto, McMaster, UBC, and Manitoba, whereas McGill states explicitly that there should be no default assumption. Formal instruments are uncommon and equally frequent (four each): the Southern Alberta Institute of Technology (SAIT) and Seneca have GenAI-specific procedures or policies, and the conduct or integrity policies of Saskatchewan Polytechnic and NAIT name AI; Western, Queen's, and Dalhousie require GenAI statements through course-outline or syllabus policy or define GenAI-related integrity departures, and McGill's Code of Student Conduct names GenAI in its academic offences. All institutions with guidance link misuse to an existing academic integrity policy, and all but one provide or expect syllabus statements (Table~\ref{tab:ordinal}). Both sectors expect program- or discipline-level adaptation (8 vs.\ 9): Red River College Polytechnic asks programs to consult industry partners and Knowledge Keepers, McMaster directs instructors to program norms set with department chairs, and Laval assigns program teams a structuring role.

\subsection{Themes and Sector Differences (RQ2)}
Disclosure, privacy, approved tools, accuracy and bias, and student AI literacy are addressed by most institutions in both sectors. Universities more often address assessment redesign (any mention 7 vs.\ 14; substantive 4 vs.\ 8), and polytechnics more often frame GenAI substantively as workforce preparation (7 vs.\ 0; any mention 9 vs.\ 8). Smaller differences, such as substantive equitable access (3 vs.\ 8) and Indigenous knowledge and data sovereignty (4 vs.\ 1), rest on few institutions. Where detectors are mentioned, universities discourage them (seven, including Waterloo, which discontinued Turnitin's AI detection in 2025), whereas polytechnics either discourage them (three) or permit them with caution (three). Rules on entering student work into GenAI tools differ: KPU requires explicit consent; Queen's and McMaster call for consent (McMaster also with an opt-out); Sheridan bars tools that sell or train on student data; and SAIT's procedure provides that students, by enrolling, are deemed to agree to the submission of their written work to an approved tool for AI-assisted feedback.

\subsection{Computing Lens (RQ3)}
Twelve institutions mention code (Fig.~\ref{fig:heat}), six in each sector, but almost always as one output type among many: in definitions of GenAI (e.g., BCIT, Fanshawe, SAIT, Sheridan, Manitoba), in lists of example uses (Algonquin; Alberta's ``code samples''), in the definition of unauthorized content generation (Queen's), or in an information-technology rule against deploying code generated by unapproved tools (Seneca). The University of Saskatchewan is the only institution with substantive (level 2) code use cases (generating code for a problem; correcting and improving code), and its non-use attestation template lists code among media; Algonquin, BCIT, and Alberta mention code uses only in passing. Manitoba notes that disciplines such as computer science may be developing their own norms, and Western's policy page points to a computer science department's rules. At least 20 of 27 institutions name a specific GenAI tool, most often Microsoft Copilot, but in their core documents none names an AI coding assistant, sets conventions for attributing generated code, addresses the licensing or provenance of generated code in student work, or gives programming-specific assessment guidance, such as code walkthroughs.

\subsection{British Columbia Vignette}
BCIT leaves decisions to faculty and discontinued AI detection in Turnitin in 2023. KPU combines a default of no use unless permitted with among the broadest values coverage in the sample, including explicit consent for student work. UBC permits GenAI in assessed work only when expressly allowed, and its definition of GenAI omits code. None of the three provides code-specific guidance, although all three offer computing programs.

\section{Discussion}

\subsection{Delegation Without Discipline-Specific Guidance} 
Institution-level guidance in both sectors assigns course-level rules to instructors and expects programs to adapt them, and two universities explicitly point to computing as a discipline with its own emerging norms. The U15's sector guide follows the same pattern: it lists code among the content generative models create and calls for support adaptable across disciplinary contexts, but offers no code-specific practices~\cite{u15guide2024}. Such delegation is defensible given disciplinary variety, but the accompanying guidance on attribution and assessment is not adapted to code (no institution provides substantive code-specific attribution or assessment guidance; Table~\ref{tab:ordinal}), so computing instructors define, course by course, what counts as acceptable help from assistants built into the editor.

\subsection{Workforce Framing and Code} 
Polytechnics justify GenAI integration as workforce preparation far more substantively than universities, yet their institution-level guidance, like that of universities, does not yet address the AI coding tools that most developers now use or plan to use \cite{stackoverflow2025}. If workforce readiness is the rationale, professional practices such as code review, licence compliance, and secure handling of credentials offer a natural basis for program-level guidance.

\subsection{Elements of Discipline-Level Guidance for Computing} 
Table~\ref{tab:template} proposes five elements, with example wording, that computing programs could add to institutional guidance. Each element addresses a computing item rarely coded as substantive (Table~\ref{tab:ordinal}), informed by prior work \cite{prather2023robots, liffiton2024codehelp}. Required syllabus statements, as at Western, Queen's, and Dalhousie, offer a vehicle for such program-level templates, which can be mapped onto the AI Assessment Scale \cite{perkins2024aias} that NAIT has adopted (document NAIT-D1 in \cite{ho2026dataset}).

\begin{table}[t]
\caption{Proposed Program-Level Statement Elements for Computing Courses}
\label{tab:template}
\centering
\footnotesize
\begin{tabular}{@{}p{1.75cm}p{6.15cm}@{}}
\toprule
Element & Example wording (to adapt per course) \\
\midrule
Assistance levels & ``AI code completion and explanation are allowed in labs; generating complete solutions to graded assignments is not.'' \\
Code attribution & ``Mark AI-assisted code with a comment naming the tool, and note briefly in the commit message how it was used (no personal or confidential data).'' \\
Tools and data & ``Use only approved assistants; never paste credentials, personal data, or code you are not permitted to share.'' \\
Licensing and security & ``You are responsible for the licence compatibility and security of any generated code you submit.'' \\
Understanding & ``Be ready to explain any submitted code in a walkthrough; your version history may be reviewed.'' \\
\bottomrule
\end{tabular}
\end{table}

\subsection{Implications for Institutions and Sector Bodies} 
Three low-cost changes follow. First, the tool approvals that most institutions already publish can state whether editor-integrated coding assistants are covered, since such tools may send code and context to external services. Second, the twelve institutions that already mention code can link those passages to program-level guidance. Third, central teaching units and sector bodies such as Polytechnics Canada and the U15 could curate and share discipline templates, such as Table~\ref{tab:template}, so that each computing program need not start from scratch.

\subsection{Limitations} 
The corpus is a snapshot of public, institution-level guidance, so an absent item means absent from public institutional guidance, not from practice; department rules, internal pages, course outlines, and content beyond the four-document cap are not captured. Extract-based codes (flagged in the dataset) may undercount, and zeros rest on reading rather than an automated keyword search. Searches can miss documents, as the re-checks showed; one university could not be retrieved; and integrity policies were not retrieved systematically, so formal instruments may be undercounted. No human coder recoded institutions independently, so single-coder bias is reduced, not eliminated. Finally, the sectors differ in size, mission, and research intensity, so differences cannot be attributed to the polytechnic model alone.

\subsection{Future Work} 
Next steps are to examine department- and course-level guidance in computing programs, as United States studies have begun to do \cite{ali2025syllabi, ganguly2026comparative}, to repeat the snapshot as guidance evolves, and to survey computing instructors and students on how delegated rules are interpreted.
\section{Conclusion}
Across 27 Canadian polytechnics and research universities with retrievable guidance, GenAI governance converges on instructor-level decisions (with differing defaults), disclosure, integrity policy, and, in most institutions, program-level adaptation, while code, the medium in which computing students learn and are assessed, appears almost always as one output type among many. Polytechnics, which most explicitly connect GenAI to workforce preparation, are well placed to lead program-level guidance for computing.

\section*{Acknowledgment}

Claude Opus 5.5 (Anthropic) was used to locate and retrieve the public documents, apply the codebook, write the analysis scripts, produce the tables and figure, and draft and revise the text of this article. The released coded dataset is labelled as AI-coded. All AI-generated content was checked for accuracy against the source documents and for potential plagiarism. All authors take responsibility for the content of this article.

\bibliographystyle{IEEEtran}
\bibliography{refs}

\end{document}